\documentclass{aastex631}
\usepackage{CJK}

\begin{document}
\begin{CJK*}{UTF8}{gbsn}

\title{In Search of the Potentially Hazardous Asteroids in the Taurid Resonant Swarm}

\correspondingauthor{Jasmine Li}
\email{li.jasmine.q@gmail.com}

\author{Jasmine Li}
\affiliation{Poolesville High School, Poolesville, MD 20837, USA}

\author[0000-0002-4838-7676]{Quanzhi Ye (叶泉志)}
\affiliation{Department of Astronomy, University of Maryland, College Park, MD 20742, USA}
\affiliation{Center for Space Physics, Boston University, 725 Commonwealth Ave, Boston, MA 02215, USA}

\author[0000-0003-4166-8704]{Denis Vida}
\affiliation{Department of Physics and Astronomy, University of Western Ontario, London, Ontario, N6A 3K7, Canada}
\affiliation{Institute for Earth and Space Exploration, University of Western Ontario, London, Ontario N6A 5B8, Canada}

\author{David L. Clark}
\affiliation{Department of Physics and Astronomy, University of Western Ontario, London, Ontario, N6A 3K7, Canada}
\affiliation{Institute for Earth and Space Exploration, University of Western Ontario, London, Ontario N6A 5B8, Canada}

\author[0000-0001-8018-5348]{Eric C. Bellm}
\affiliation{DIRAC Institute, Department of Astronomy, University of Washington, 3910 15th Avenue NE, Seattle, WA 98195, USA}

\author[0000-0002-5884-7867]{Richard Dekany}
\affiliation{Caltech Optical Observatories, California Institute of Technology, Pasadena, CA 91125, USA}

\author[0000-0002-3168-0139]{Matthew J. Graham}
\affiliation{Division of Physics, Mathematics, and Astronomy, California Institute of Technology, Pasadena, CA 91125, USA}

\author[0000-0002-8532-9395]{Frank J. Masci}
\affiliation{IPAC, California Institute of Technology, 1200 E. California Blvd, Pasadena, CA 91125, USA}

\author[0000-0003-1227-3738]{Josiah Purdum}
\affiliation{Caltech Optical Observatories, California Institute of Technology, Pasadena, CA 91125, USA}

\author[0000-0001-8861-3052]{Benjamin Racine}
\affiliation{Aix Marseille Univ, CNRS/IN2P3, CPPM, Marseille, France}

\author[0000-0002-9998-6732]{Avery Wold}
\affiliation{IPAC, California Institute of Technology, 1200 E. California Blvd, Pasadena, CA 91125, USA}

\begin{abstract}

The Taurid Complex is a large interplanetary system that contains comet 2P/Encke, several meteoroid streams, and possibly a number of near-Earth asteroids. The size and nature of the system has led to the speculation that it was formed through a large-scale cometary breakup. Numerical investigations have suggested that planetary dynamics can create a resonant region with a large number of objects concentrated in a small segment of the orbit, known as the Taurid swarm, which approaches the Earth in certain years and provides favorable conditions to study the Taurid Complex. Recent meteor observations confirmed the existence of the swarm for mm- to m-sized objects. Here we present a dedicated telescopic search for potentially hazardous asteroids and other macroscopic objects in the Taurid swarm using the Zwicky Transient Facility survey. We determine from our non-detection that there are no more than 9--14 $H\leq24$ (equivalent to a diameter of $D\gtrsim100$~m) objects in the swarm, suggesting that the Encke--Taurid progenitor was $\sim10$~km in size. A progenitor of such a size is compatible with the prediction of state-of-the-art Solar System dynamical models, which expects $\sim0.1$ $D>10$~km objects on Encke-like orbits at any given time.

\end{abstract}


\section{Introduction} \label{sec:intro}

The Taurid Complex comprises comet 2P/Encke, several prominent meteoroid streams such as the Northern and Southern Taurids, and daytime showers Beta Taurids and Zeta Perseids \citep[cf.][]{1987PAICz..67..167P, 2006CoSka..36..103P}. A number of near-Earth asteroids (NEAs) were also suggested to be a part of the Complex \citep[cf.][]{1993MNRAS.264...93A, 2008MNRAS.386.1436B, 2021MNRAS.507.2568E}, some of which are potentially hazardous (defined as having a minimum orbit intersection distance with Earth of $<0.05$~au and an absolute magnitude of $H\leq22$). The complicated structure of the Taurid Complex presents an excellent case to study the origin and evolution of meteoroid streams. Specifically, the unusual abundance of large, meter-class meteoroids within the meteoroid stream led to the controversial speculation that the Taurid Complex was the relic of a very large, possibly 100-km-class comet that disrupted 10--20 kyr ago \citep{1984MNRAS.211..953C}. 

While recent advances in the study of near-Earth objects (NEOs) have now rendered a 100-km-class progenitor for the Taurid Complex exceedingly unlikely, numerical investigations of the comet break-up theory showed that debris produced by such a break-up would tend to be trapped within a 7:2 orbital resonance with Jupiter, creating a ``swarm'' \citep{1993QJRAS..34..481A}. Their model predicts increased fireball activities and approaching Taurid-like asteroids in certain years when the Taurid swarm approaches the Earth \citep{1994VA.....38....1A}, which have been confirmed by meteor data \citep[e.g.][]{1998MNRAS.297...23A, 2017MNRAS.469.2077O}. \citet{2017A&A...605A..68S} and \citet{2023LPICo2851.2066S} unambiguously reported several Taurid fireball outbursts occurring at the time the Earth's passage through the predicted Taurid swarm. Accurate orbital measurements of mm- to m-sized Taurids showed that they were tightly clustered in a narrow range of semi-major axes and mean anomalies corresponding to the resonant region. However, the enhancement of the larger, asteroidal counterparts has not been unambiguously detected. Two Taurid-like asteroids were found serendipitously during the 2015 Taurid swarm encounter \citep{2016MNRAS.461..674O} but the statistical significance of this discovery was unclear. A dedicated campaign was proposed to search for Taurid-like asteroids during the more favorable 2019 encounter\citep{2019MNRAS.487L..35C} , but the attempt fell victim to the weather and technical issues.

In 2022, the Earth would once again pass close to the center of the Taurid swarm, albeit not as favorable as either of the 2015 and 2019 encounters. Here we present a dedicated search for the hypothetical Taurid swarm asteroids using the Zwicky Transient Facility (ZTF) Camera.

\section{Observations}

The ZTF Camera is a dedicated mosaic camera installed on the Palomar Observatory 1.2 m Oschin Schmidt telescope with a field-of-view of 47 deg$^2$ \citep{2020PASP..132c8001D}. The ZTF survey began in 2018 and conducts repeat imaging of the visible sky every two to three nights \citep{2019PASP..131a8002B, 2019PASP..131g8001G}. With a typical survey exposure of 30 s, ZTF can observe an area of 3760 deg$^2$ every hour down to a $5\sigma$ limit of $r=20.7$. The use of a large-aperture, wide-field camera such as the ZTF Camera is necessary, since the search area of the Taurid swarm asteroids typically spans over many tens of degrees in the sky, with a 100-meter-diameter asteroid having $r\gtrsim20$ magnitude \citep[][Fig. 3--7]{2019MNRAS.487L..35C}.

We followed the procedure described in \citep{2019MNRAS.487L..35C} to guide the search. In brief, we generated 10,000 synthetic Taurid asteroids based on the high-precision fireball orbits reported by \citet{2017A&A...605A..68S} and forced them inside the semimajor axis range appropriate for the 7:2 resonance zone. We then rotated the mean anomalies of the particles to $\pm40^\circ$ to cover the estimated extent of the Taurids swarm \citep[cf.][]{1991MNRAS.251..632S, 1993QJRAS..34..481A}. While the value of $\pm 40^\circ$ was originally an empirical estimate, fireball observations in recent years are mostly in line with this estimate, with the largest outlier at $48^\circ$ \citep{2023LPICo2851.2066S}. To allow for brightness estimation for planning purposes, all these asteroids were assumed to have an absolute magnitude $H=24$, equivalent to a diameter of 100 m assuming a comet-like geometric albedo of 0.04.

\begin{figure*}
    \centering
    \includegraphics[width=1\linewidth]{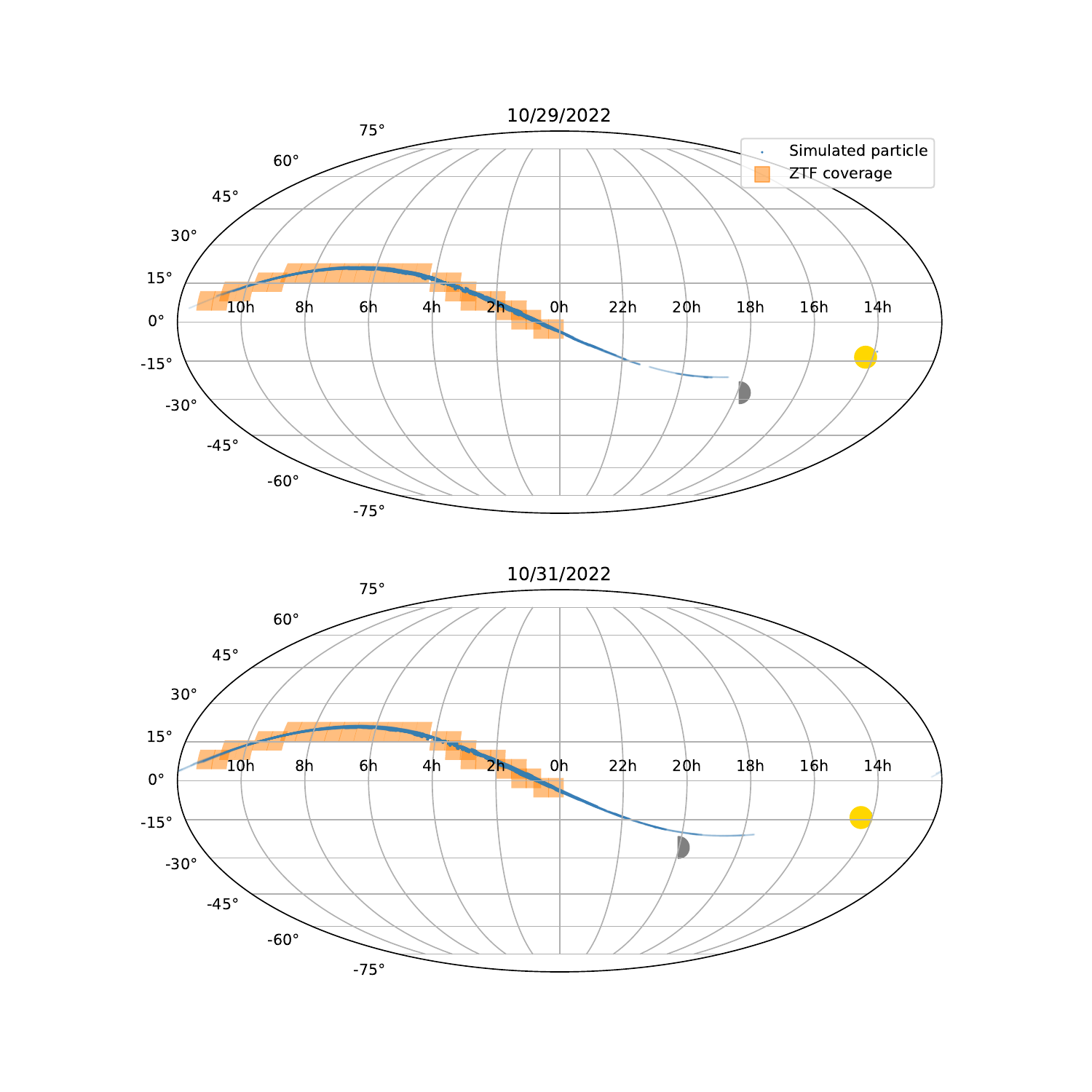}
    \caption{Simulated Taurid swarm asteroids with $V<21$ (blue dots) and ZTF footprint (filled orange squares) on Oct 29 and 31, 2022. The ZTF search covers 99\% of the simulated asteroids. The locations of the Sun and the Moon are marked with a yellow circle and a gray half-filled circle.}
    \label{fig:cov}
\end{figure*}

We conducted observations with ZTF on 2022 October 29 and 31 to cover the predicted positions of the simulated asteroids. After excluding particles that are fainter than $V=21$ and areas too close to the Sun and/or the Moon ($<45^\circ$), the entire search area covers 99\% of the $V<21$ particles, and can be covered by 33 ZTF pointings. Each pointing was observed twice in order to reveal moving objects. The on-sky positions of the simulated asteroids as well as the survey footprint are shown in Figure~\ref{fig:cov}.

Images were automatically processed using the ZTF Science Data System \citep{2019PASP..131a8003M} which produced bias and flat-corrected images, background-subtracted difference images, and source catalogs. The images were photometrically calibrated using the Pan-STARRS DR1 catalog \citep{2020ApJS..251....6M}. As part of the data processing procedure, images were searched for moving objects, including trailed fast-moving objects \citep{2019MNRAS.486.4158D, 2019PASP..131g8002Y} and slower objects, but no candidates matching the predicted motions of the Taurids were found.

\section{Analysis}

The search done by the regular data processing pipeline is by no means exhaustive, as the identification of new objects requires multiple sightings/detections of the same object (at least two for streaked objects, at least four for non-streaked objects). Thus, we conduct a targeted, more sensitive search by focusing on the parameter space covered by the synthetic Taurid asteroids.

The apparent speeds of the Taurid asteroids strongly depend on where we are looking. Most of the search fields target asteroids that are close to the Earth and thus would streak in ZTF images; only a few fields target asteroids that are far enough to show up as point sources in ZTF images. Hence, we use two different approaches to search the image, guided by the expected apparent speed of the simulated asteroids in the image.

For the images that are expected to contain slow-moving Taurid asteroids, we focus on detecting point-source-like objects. The regular data processing pipeline only detects objects on the difference images down to a signal-to-noise ($S/N$) level of 5, thus we will miss fainter objects as well as objects involved with background stars. To address these issues, we reran object detection on the difference images using the Python library for Source Extraction and Photometry version 1.2.1 \citep[SEP;][]{1996A&AS..117..393B, Barbary2016}, with threshold pixel value for detection set to 2.0 and other parameters set as default. We then found that most of the detections were contributed by subtraction artifacts of background stars, hence we removed sources that were within 20 pixels of any source identified in the companion image, leaving 20--30 sources per image. These remaining sources were then visually examined to reject junk sources, which typically left less than ten sources per field. We then use the SkyBoT tool \citep{2006ASPC..351..367B} to check these sources against known asteroids. All of the remaining sources turn out to be known asteroids.

For images that are expected to contain streaked Taurid asteroids, we extract raw detections made by the ZSTREAKS pipeline \citep{2019MNRAS.486.4158D, 2019PASP..131g8002Y} before the machine-learning filtering. After sorting the detected streaks by sky field, streaks that matched the broad range of direction and motion of the simulated asteroids in the fields of interest were then extracted for further analysis, leaving about 25,000 streaks from each night. We then visually inspected the thumbnails of these streaks generated by ZSTREAKS. The vast majority of these detections were false positives or satellite flashes, but our analysis revealed one real asteroid, 2022 UL$_{16}$, discovered by the Asteroid Terrestrial-impact Last Alert System (ATLAS) with pre-discovery detections by the NEOWISE mission and our ZTF campaign. A closer examination found that 2022 UL$_{16}$'s orbit is noticeably different from that of Taurid asteroids ($a=2.05$, $i=21^\circ$ versus $a\sim2.2$, $i\sim10^\circ$ of Taurids), and thus does not belong to the Taurid swarm.

\section{Discussion}

Although no Taurid swarm asteroid was found, our non-detection result can still be used to constrain the maximum density of the swarm. It is important to note that the trailing of potential Taurid asteroids results in a loss of sensitivity, which must be accounted for when estimating detection efficiency. We first calculate the average particle motion rate for each field and verify that the speed dispersion within a given field is sufficiently small (the dispersion of the corresponding displacement being on the order of a pixel or less). Subsequently, using the established relationship between average motion rate and ZSTREAK's detection limit \citep{2019PASP..131g8002Y}, we determine the detection limit for each field and compare them against the predicted brightness of the simulated particles. Out of the fields of interest, the motion rates were actually quite similar, resulting in the detection limit for each field being $V\sim20$. We found that 1,088 and 718 particles out of the simulated 10,000 particles would be visible on 2022 October 29 and 31 respectively, with the lower number on the latter date probably due to the higher particle motion rate. Given that the simulation assumes a flat $H$ distribution of $H=24.0$ (equivalent to 100-meter-class asteroids with comet-like albedo), we conclude from our non-detection that there are no more than $10000/(718~\mathrm{to}~1088)=$ 9--14 objects of $H\leq24$ within the swarm.

Among the near-Earth asteroids that might be associated with the Taurid Complex, a handful were serendipitously discovered during close encounters of the Taurid swarm: 2005 TB$_{15}$, 2005 TF$_{50}$, 2005 UR, and (452639) 2005 UY$_{6}$ were found during the 2005 encounter \citep{2011MmSAI..82..310J}, and 2015 TX$_{24}$ was found during the 2015 encounter \citep{2017A&A...605A..68S}. 2005 UY$_{6}$ is the largest object in this sample with a diameter of $2.2\pm1.1$~km \citep{2017AJ....154..168M}; the other three asteroids are sub-km in diameter. Additionally, \citet{2021MNRAS.507.2568E} identified 15 asteroids found in 2019--20 that may be associated with the swarm, 7 of which had $H<24.0$ with the largest one at $H=20.0$ (2020 BC$_{8}$). However, the same authors also found that of all these asteroids, only 2005 TF$_{50}$, 2005 UR and 2015 TX$_{24}$ exhibit orbital convergence to one another or 2P/Encke in recent history (20~kyr). The current catalog of NEOs that have minimum orbit intersection distances (MOID) with Earth of $<0.05$~au and $H\leq24.0$ is only $\sim15\%$ complete \citep{2023AJ....166...55N}. With a total of 3 high-confidence objects, there should be no more than $3 / 15\% \sim20$ Taurid swarm asteroids with $H\leq24.0$, and this may be an overestimation given that the completion of Taurid swarm asteroids should be higher than normal NEOs given their periodical approaches to the Earth. In all, the past observations are broadly in line with our estimate of the number of Taurid swarm asteroids.

Next, we attempted to constrain the mass of the swarm, which subsequently provided clues to the size of the proto-Encke object. The mass of the swarm depends on the poorly understood size distribution of the swarm objects, but all major small body populations are ``top-heavy'' -- i.e. the total mass is primarily contributed by the most massive objects, and we have no reason to believe that the Taurid Complex would behave otherwise. The largest swarm object known to date is 2005 TF$_{50}$, which has $H=20.3$. Taking a survey completeness of $H=20$ NEOs to be $20\%$ \citep{2023AJ....166...55N}, we expect around five $H=20$ objects in the swarm. By simply adding up these five objects, we estimate that the size of the direct progenitor is at the order of $\sim10$~km. Since planetary dynamics has likely driven most of the mass in the Taurid Complex besides 2P/Encke \citep[which is 5~km in diameter; ][]{2004come.book..223L} into the Taurid swarm, we postulate that the proto-Encke object may not be much larger than $\sim10$~km. (The Taurid meteoroid stream has a mass of $1\times10^{13}$~kg, cf. \citealt{1994A&A...287..990J}, equivalent to a $D\sim3$~km body, and thus does not contribute significantly to the total mass of the Taurid Complex.) Compared to a $\sim100$-km-class proto-Encke object proposed by the giant comet hypothesis, a $\sim 10$~km proto-Encke object is more compatible with contemporary observational-model constraints: through numerical investigations, \citet{2006Icar..182..161L} determined that there would be $\sim12$ objects (including both the active and inactive ones) with $D>1$~km in Encke-like orbits. Taking the size distribution of scattering trans-Neptunian objects \citep[the primary source of short-period comets][]{2017ApJ...845...27N}, this would imply a total of $\sim0.1$ $D>10$~km objects and $\sim0.001$ $D>100$~km objects on Encke-like orbits at any given time, with the numbers good to order-of-magnitude level. A proto-Encke object of $D\sim100$~km is statistically unlikely to appear and would produce a Taurid Complex that is several orders of magnitude more massive than supported by contemporary observations.

\section{Conclusion}

Through a dedicated search using the Zwicky Transient Facility survey, we determined that there are no more than 9--14 objects of $H\leq24$ ($D\gtrsim100$~m) in the Taurid swarm, a dynamically enhanced concentration of fragments and dust of the Taurid Complex. This suggests the progenitor that produced the Complex is probably not much larger than $\sim10$~km in size. A progenitor of such a size is also consistent, on an order-of-magnitude level, with the expectations from the state-of-the-art planetary dynamical models.

\begin{acknowledgments}
This work is supported by NASA program 80NSSC22K0772. DV was supported in part by NASA Cooperative Agreement 80NSSC21M0073 and by the Natural Sciences and Engineering Research Council of Canada. Based on observations obtained with the Samuel Oschin Telescope 48-inch and the 60-inch Telescope at the Palomar Observatory as part of the Zwicky Transient Facility project. ZTF is supported by the National Science Foundation under Grant No. AST-2034437 and a collaboration including Caltech, IPAC, the Weizmann Institute of Science, the Oskar Klein Center at Stockholm University, the University of Maryland, Deutsches Elektronen-Synchrotron and Humboldt University, the TANGO Consortium of Taiwan, the University of Wisconsin at Milwaukee, Trinity College Dublin, Lawrence Livermore National Laboratories, IN2P3, University of Warwick, Ruhr University Bochum, Cornell University, and Northwestern University. Operations are conducted by COO, IPAC, and UW.
\end{acknowledgments}

\vspace{5mm}
\facilities{PO:1.2m}

\software{SkyBoT \citep{2006ASPC..351..367B}, 
          Python library for Source Extraction and Photometry \citep{1996A&AS..117..393B, Barbary2016}
          }

\bibliography{ref}{}
\bibliographystyle{aasjournal}

\end{CJK*}
\end{document}